# Design and Implementation of Smart Cooking based on Amazon Echo

Lin Xiaoguang [1, 2, 3], Yang Yong[3], Zhang Ju[1, 3]

[1] University of Chinese Academy of Sciences, Beijing, China
`lxg@cigit.ac.cn`
[2] Chengdu Computer Applications Institute, Chinese Academy of Sciences, Chengdu, China
[3] Chongqing Institute of Green and Intelligent Technology, Chinese Academy of Sciences, Chongqing, China

## ABSTRACT

*Smart cooking based on Amazon Echo uses the internet of things and cloud computing to assist in cooking food. People may speak to Amazon Echo during the cooking in order to get the information and situation of the cooking. Amazon Echo recognizes what people say, then transfers the information to the cloud services, and speaks to people the results that cloud services make by querying the embedded cooking knowledge and achieving the information of intelligent kitchen devices online. An intelligent food thermometer and its mobile application are well-designed and implemented to monitor the temperature of cooking food.*

## 1. INTRODUCTION

A smart home is an application of ubiquitous computing in which the home environment is monitored by ambient intelligence to provide context-aware services and facilitate remote home control [1]. With the popularity of the internet of things, many domestic appliances could process some intelligent computing and access to the internet with WIFI. People may monitor and control these appliances using smartphone anytime and anywhere. The one could check the indoor temperature and open the air-conditioner to cool down to the comfortable temperature by operating the specific apps in his smartphone before he get home.

Smart cooking is one of the typical smart home scenes. Cooking has become a basic necessity for human beings, since food is one of the basic human needs [2]. There are a variety of kitchen appliances, like the oven, the induction cooker, the electric stove, the food temperature, etc. Cooking food usually need to use some different kitchen appliances at the same time. Especially, in the Far East, people cook rice with electric stoves, heat ingredients with microwave ovens, and stew soup with induction cookers. How to integrate these kitchen appliances effectively is the key task in smart cooking. Cooking techniques vary widely across the world, from grilling food over an open fire to using electric stoves. However, monitoring the temperature of the cooking food is the core issues of common concern. Usually, the cooking is boring, and the cook may spend a lot of time to heat food to target temperature. And many cooks are not rich in cooking knowledge, even they don't know when the beef has been cooked medium well. So, a well-designed smart cooking system should be helpful for these cooks.

The cooks, usually the housewives, may spend a lot of time in the kitchen to prepare the breakfast, the lunch, the dinner, and the supper. They often feel bored and lonely when they are cooking. Thus, they will choose listening the music or the radio to amuse the cooking. Nowadays, many smart speakers, like Amazon Echo series, and Google Home, are invented with intelligent human-computer interaction using the internet of things and cloud computing. The Amazon Echo is introduced in this paper to assist in cooking. People may speak to Amazon Echo during the cooking in order to get the information and situation of the cooking. Amazon Echo recognizes what people say, then transfers the information to the cloud services, and speaks to people the results that cloud services make by querying the embedded cooking knowledge and achieving the information of intelligent kitchen appliances online.

This paper introduces the design and implementation of a well-designed smart cooking system based on Amazon Echo. The remainder of this paper is organized as follows. Section 2 shows the architecture of the smart cooking. Section 3 introduces the smart speaker and how to implement an Amazon Alex Skill. Section 4 shows a customized intelligent food thermometer for smart cooking. Section 5 shows the design and implementation of the cloud services of smart cooking based on Amazon Echo.

## 2. ARCHITECTURE

The smart cooking based on Amazon Echo could integrate a variety of kitchen devices using cloud computing, as shown in Figure 1.

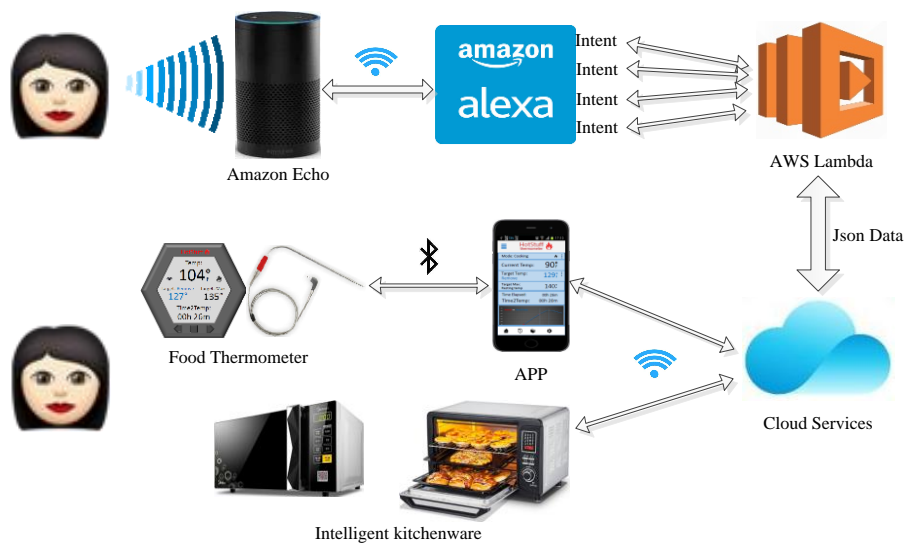

Figure 1. Smart Cooking Architecture based on Amazon Echo

Amazon Echo is the core hardware of the whole smart cooking. Amazon Echo could be awaken by some specific "wake words" (like "Alexa", "Echo", etc.). Then, Amazon Echo will catch the sound what the person say, for example "what is the current temperature of my food", and convert sound signals into some digital information successively. On the other side, Amazon Echo will receive and then speak out the result sentences returned from services, for example "the current temperature of you food is 120 degrees Fahrenheit".

Amazon Alexa Voice Service online will receive the information that Amazon Echo send through WIFI, and recognize the information by the inner speech recognition program. And then, one of the pre-designed Intents should be matched according to the identified key words. Different intent corresponds to different service of Amazon Web Services (AWS for short).

AWS Lambda will execute the code written in Node.js with passed parameters from Amazon Alexa, and call the customized cloud services remotely when the necessary information is needed. And then, AWS Lambda will send back the result sentences to Amazon Alexa after completion of the execution.

Cloud Services keep the connections of intelligent kitchenware and gather the condition information into embedded database in real time. Cloud Services will encapsulate the information inquired from the database in JSON format, and transmit the JSON data to AWS Lambda. Cloud Services could directly connect the intelligent kitchenware equipped WIFI, and indirectly connect the other kitchenware equipped Bluetooth through accessory APPs.

Food Thermometer with an insertable probe and its well-designed mobile application are essential to monitor the temperature of the cooked food, especially the meat, and predict time remaining to specific doneness.

## 3. SMART SPEAKER AND AMAZON ECHO

A smart speaker is a type of wireless speaker and voice command device with an integrated virtual assistant (artificial intelligence) that offers interactive actions and hands-free activation with the help of one or several "wake words". A smart speaker is a smart device that utilizes WIFI, Bluetooth and other wireless protocol standards to extend usage beyond audio playback, such as to control home automation devices. Each can have its own designated interface and features in-house, usually launched or controlled via application or home automation software [4].

Amazon Echo is a branch of smart speaker developed by Amazon.com. The devices connected to the voice-controlled intelligent personal assistant service Alexa, which responds to the name "Alexa". This "wake word" can be changed by the user to "Echo" or other words. The device is capable of voice interaction, music playback, making to-do lists, setting alarms, streaming podcasts, playing audiobooks, and providing weather, traffic and other real-time information. It can also control several smart devices acting as a home automation hub [5].

Amazon Alexa Voice Service is a virtual assistant developed by Amazon.com. It is capable of voice interaction, music playback, making to-do lists, setting alarms, streaming podcasts, playing audiobooks, and providing weather, traffic, sports, and other real-time information, such as news. Users are able to extend the Alexa capabilities by installing "skills" (additional functionality developed by third-party vendors, in other settings more commonly called apps such as weather programs and audio features). Amazon allows developers to build and publish skills for Alexa Skills Kit. Most skills run code almost entirely in the cloud, using Amazon's AWS Lambda services [6].

AWS Lambda is an event-driven, serverless computing platform provided by Amazon.com as a part of the Amazon Web Services. It is a computing service that runs code in response to events and automatically manages the computing resources required by that code. The purpose of Lambda, is to simplify building smaller, on-demand applications that are responsive to events and new information. Node.js, Python, Java, Go and C# through .NET Core are all officially supported, and other languages can be supported via call-outs. AWS Lambda can also be automatically provision back-end services triggered by custom HTTP requests, and "spin down" such services when not in use, to save resource [7].

An Amazon Alexa custom skill is shown in Figure 2. Customer may ask a question or give a command; Alexa identifies the skill's name, analyses and understands the user's request, then sends the service a structured representation of the user's request; service processes the request and returns a text; Alexa converts the returned text to speech and streams it to Echo; Customer hear the response from Alexa's voice.

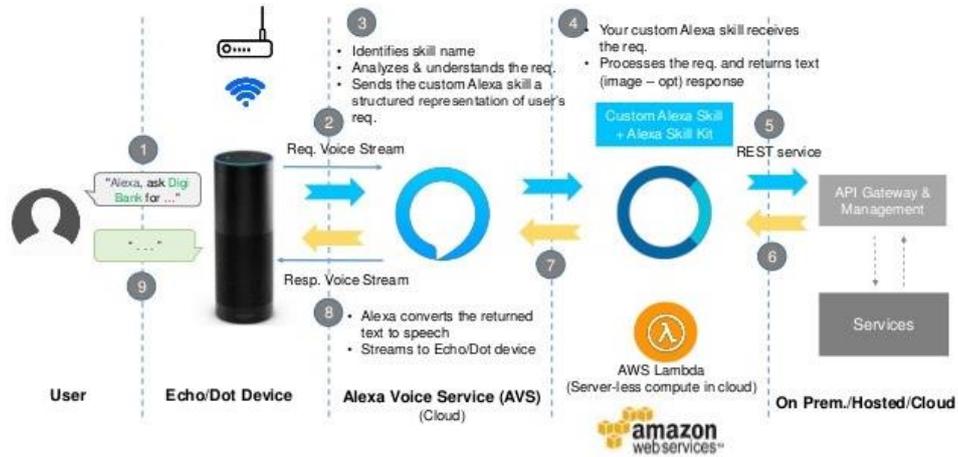

Figure 2. Amazon Alexa Custom Skill – Reference Architecture

## 4. INTELLIGENT FOOD THERMOMETER

Food thermometers such as meat thermometers have been used to help provide more accurate and consistent cooking results. The use of a meat thermometer, for example, can provide a visual indication on whether the meat is still undercooked or overcooked. However, these conventional types of food thermometers provide a passive indication of temperature and generally rely on the cook to remember to check the temperature, and may not provide sufficiently accurate information during cooking, such as a completion time for specific doneness, when to adjust a temperature, when to start or finish a particular cooking stage such as searing, or how long to let the food rest after removing it from heat.

As an important component of smart cooking, an intelligent food thermometer and its mobile application is designed and implemented in this paper as Figure 3.

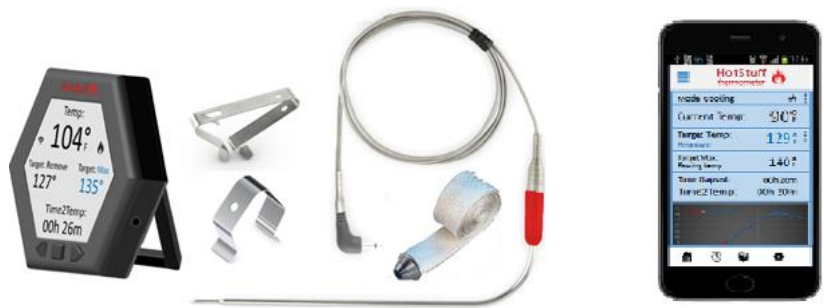

Figure 3. An Intelligent Food Thermometer and its APP

The main firmware of the intelligent food thermometer contains a development board, a LCD display panel and three control buttons. The development board is embedded with temperature sensing module, Bluetooth communication module, digital to analog conversion module and data processing module. The temperature sensing module obtains the temperature signal via a high temperature 4.5-inch curved penetration probe. The LCD display panel in the front of the firmware could display some important number information, such as the current temperature value, the target max temperature value set by cooks, and the elapsed time value. The control buttons could be used to set up or down the target max temperature and start the cooking time.

The main firmware of the intelligent food thermometer includes the following accessories: pot clip for submerging probe into pots; grate clip to elevate probe above oven or grill grates; 2 foot 1200°F continuous exposure braided fiberglass probe-wire sleeve.

The accessional APP of the intelligent food thermometer connected with the main firmware via Bluetooth displays the consistent number information synchronously. Some additional knowledge is saved in the embedded SQLite database of the APP, such as the food information, the ingredient information, and the meat doneness shown in Table 1. The APP uses a real-time scatter diagram to visualize the change of the temperature transmitted from the thermometer.

Table 1. Meat Doneness (from USDA guidelines)

| Doneness | Serving temperature | Description |
| --- | --- | --- |
| Beef, lamb, veal steaks, chops, roasts & duck breasts (USDA recommended minimum: 145°F plus 3 minutes rest) | | |
| Extra rare | 110-120°F | Bright purple-red center, cool, stringy, tender, slippery, slightly juicy |
| Rare | 120-130°F | Dark red center, warm, tender to mildly firm, juicy |
| Medium rare | 130-135°F | Light red center, warm, mildly firm, very juicy |
| Medium | 135-145°F | Pink center, firm, slightly juicy |
| Medium well | 145-155°F | Tan with slight pink center, firm and slightly fibrous, some juice |
| Well done | More than 155°F | Tan to brown center, no pink, chewy, little juice |
| Pork & veal steaks, chops & roasts (USDA recommended minimum: 145°F plus 3 minutes rest) | | |
| Raw | Less than 120°F | Bright pink center, cool, stringy, slightly juicy |
| Rare | 120-130°F | Pale pink center, warm, tender, very juicy |
| Medium rare | 130-135°F | Cream colored with a slight pink tinge, tender, juicy |
| Medium | 135-145°F | Cream colored, firm, slightly pink juices |
| Medium well | 145-155°F | Cream colored, firm, clear juices |
| Well done | More than 155°F | Cream colored, tough, clear juices |
| Turkey & chicken, whole or ground (USDA recommended minimum: 165°F) | | |
| Safe and moist | 165°F | Cream colored, tender, clear juices |
| Fish (USDA recommended minimum: 145°F) | | |
| Rare | 125°F | Similar to the raw meat in color, just a bit paler |
| Medium | 135°F | Slightly translucent meat, flakes easily |
| Well done | 145°F | Opaque, pearly meat |

The most helpful function of the APP is dynamically predicting and showing food's completion time to specific doneness. The remaining completion time could be calculated by the following formula:

$$Time_{remaining\ completion} = \triangle_{time} / \triangle_{temp} \times (Temp_{target} - Temp_{current}).$$

The APP maintains the communication with cloud services via WIFI, through receiving Restful http request from cloud services and sending the response back to cloud services.

## 5. CLOUD SERVICES

Cloud services in the smart cooking may be built in the public cloud infrastructure, such as Amazon EC, Ali Cloud, Tencent Cloud, and etc. Cloud services maintain the states of all the connected kitchenware, receive the querying request from AWS Lambda, and send the result information back to AWS Lambda. All requests are Restful (representational state transfer) and all transferred data is formatted in JSON.

The connection between the cloud and the APP introduced in Part 4 will be established and then maintained when the APP is opened. Some other intelligent kitchenware satisfied the communication standard will be also connected by the cloud via WIFI.

As one of the cooking knowledge base, a cooking ontology is designed for the smart cooking by protégé in cloud services. The cooking ontology includes some necessary knowledge, such as the food (meat and vegetables), the indigent, the recipes, and some cooking notes. The meat doneness shown in Table 1 is also structuralized in owl format.

## 6. IMPLEMENTATION

Many different types of components are integrated in the smart cooking based on Amazon Echo. Thus, some configuration and programming are necessary. Firstly, use cases of Amazon Echo should be well-designed, including sentences recognized by Amazon Alexa and corresponding expected sentences spoken out by Amazon Echo. Table 2 shows parts of Amazon Alexa Skill use cases for the intelligent Thermometer, such as querying current temperature of the cooking food, setting target temperature of the cooking food, querying the completion time, and setting alarm.

Table 2. Parts of Amazon Alexa Skill Use Cases for the Intelligent Thermometer.

| Intent | Sentences | Expected Reply |
| --- | --- | --- |
| Query Current Temperature | What's the temperature of my food? | Your food is currently at ** degrees Fahrenheit. |
| | How hot is my food? | |
| | What's the current temperature of my food? | |
| Set Target Temperature | Set the target temperature to ** degrees. | Ok, your Target Temperature has been set to ** degrees. |
| | Set thermometer to ** degrees. | |
| | Tell thermometer to set the temperature alarm for ** degrees. | |
| Query Completion Time | When will my food be ready? | Your thermometer predicts that the time-to-temperature is xxx minutes. |
| | How long until my food is done? | |
| | Is may food ready? | |
| Set alarm | Notify me when my food is done. | Ok, your temperature alarm will ring unit as well |
| | Set an alarm for when my food is ** degrees. | |
| | Let me know when my food is ** degrees. | |

Secondly, configure Alexa skills using Alexa Skills Kit in a JSON file. Parts of the configuration shown as following.

```
{
  "name": "CurrentTempIntent ",
  "samples": [
     "what's the temperature",
     "how hot is my {Food_ct}",
     "what is the {place_ct} temperature of my {Food_ct}"
  ]
}
{
  "name": "SetTargetTempIntent",
  "samples": [
     "set the temperature {Warn_stt} for {Temp_stt} degrees",
     "set thermometer to {Temp_stt} degrees",
     "tell thermometer to set the temperature alarm for {Temp_stt} degrees"
  ]
}
{
  "name": "CookTimeIntent",
  "samples": [
     "Is my {Food_cti} {Complete_cti}",
     "how long until my {Food_cti} is {Complete_cti}",
     "when will my {Food_cti} be {Complete_cti}"
  ]
}
```

```json
{
    "name": "SetTargetAlarmIntent",
    "samples": [
      "notify me when my {Food_stai} is {Complete_cti}",
      "set an {alert_stai} for when my food is {Complete_stai}",
      "notify me when my food is {Complete_stai}"
    ]
}
```

Thirdly, some programming shall be written in AWS Lambda using Node.js (version 8.10) to process the communication with cloud services, sending the RESTful request to cloud services and receiving the response from cloud services. The programming code of querying current temperature intent, one of intents designed in Table 2, is shown as following.

```
const CurrentTempIntentHandler = {
  canHandle(handlerInput) {
    const request = handlerInput.requestEnvelope.request;
    return request.type === 'LaunchRequest' || (request.type === 'IntentRequest' && request.intent.name === 'CurrentTempIntent');
  },

  handle(handlerInput) {
    try{
        let request = handlerInput.requestEnvelope.request;
        let intentName = request.intent.name;
        let sessionId = handlerInput.requestEnvelope.session.sessionId;
        let slotValue = request.intent.slots.Food_ct.value;
        let tokenId = handlerInput.requestEnvelope.session.user.accessToken;
        var options = {
            host: '140.143.237.143',
            port: 80,
            path: '/NewHotStuff/Aimtemp?token='+tokenId,
            method: 'GET'
        };
        return new Promise((resolve, reject) => {
            httpGet(options).then((response) => {
                console.info(response);
                let responseData = JSON.parse(response);
                let speechText = `${responseData.message}`;
                resolve(handlerInput.responseBuilder.speak(speechText).reprompt(`${repromptText}`)
                            .withShould0EndSession(false).getResponse());
            }).catch((error) => {
                resolve(handlerInput.responseBuilder.speak('Internet error.').reprompt(`${repromptText}`).getResponse());
            });
        });
    }catch(error){
        return handlerInput.responseBuilder.speak('Internal error. ').getResponse();
    }
  },
};
```

Next, RESTful cloud services should be implemented in many programming languages, like JAVA, PHP, C#.net, and Python. The common interfaces of could services include adding data into database or querying data from database using SQL, and querying knowledge from the cooking ontology using Jena and SPARQL. The implementation of communication with the intelligent thermometer and other intelligent kitchenware is shown in part 4.

We have implemented the code, and we plan to share some useful programs on a GitHub repository.

## 7. CONCLUSION

A lot of intelligent devices are integrated in the smart cooking based on Amazon Echo, including Amazon Echo, intelligent thermometer, and intelligent kitchenware. Many advanced technologies are used to implement the smart cooking based on Amazon Echo, such as cloud computing, the things of internet, speech recognition, natural language processing, knowledge base, RESTful web services, smartphone APP and mobile internet.

The smart cooking based on Amazon Echo provides some basic use cases for cook as a result of Amazon Echo's limited functions. With the improving of Amazon Echo and Amazon Alexa, the smart cooking will match more intelligent kitchenware and provides more complicated use cases.

## ACKNOWLEDGEMENTS

This paper is funded by Chongqing S&T Foundation Project in China. The project number is cstc2015ptfw-ggfw120002.

## REFERENCES

[1] Alam M R, Reaz M B, Ali M A, (2012) "A Review of Smart Homes—Past, Present, and Future", systems man and cybernetics, Vol 42, No. 6, pp1190-1203.

[2] Alif Ahmad Syamsudduha, Dyah Pratiw, etc, (2013) "Future Smart Cooking Machine System Design", TELKOMNIKA, Vol.11, No.4, pp827~834

[3] Hashimoto Atsushi, Mori Naoyuki, etc, (2008) "Smart Kitchen: A User Centric Cooking Support System", Proceedings of IPMU'08, pp848-854.

[4] https://en.wikipedia.org/wiki/Smart_speaker.

[5] https://en.wikipedia.org/wiki/Amazon_ Echo.

[6] https://en.wikipedia.org/wiki/Amazon_ Alexa.

[7] https://en.wikipedia.org/wiki/AWS_Lambda.

[8] Beth M Sheppard, (2017) "Theological Librarian vs. Machine: Taking on the Amazon Alexa Show (with Some Reflections on the Future of the Profession)", Theological Librarianship, Vol 10, issue 1, pp8-23.

**Authors**

Lin Xiaoguang

Lin Xiaoguang is a Ph.D. candidate in University of Chinese Academy of Sciences.

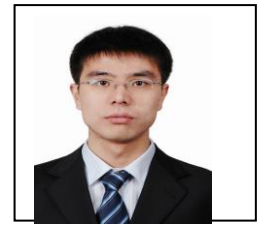